\documentclass[aps,prl,twocolumn,reprint,superscriptaddress,citeautoscript]{revtex4-2}

\usepackage{physics}
\usepackage{graphicx}
\usepackage{amsmath,amssymb}
\usepackage[normalem]{ulem}
\usepackage{bm}
\usepackage[whole]{bxcjkjatype}

\usepackage{xcolor}

\usepackage{hyperref}
\hypersetup{
	citecolor = blue,
	colorlinks = true,
	urlcolor = blue
}

\begin{document}
	
\title{
What is a proper definition of spin current? \\
--Lessons from the Kane-Mele Model}

\author{Tomohiro Tamaya}
\email{tamaya@g.ecc.u-tokyo.ac.jp}
\affiliation{Institute for Solid State Physics, University of Tokyo, Kashiwa, 277-8581, Japan}
\affiliation{JST, PRESTO, 4-1-8 Honcho, Kawaguchi, Saitama, 332-0012, Japan}

\author{Takeo Kato}
\affiliation{Institute for Solid State Physics, University of Tokyo, Kashiwa, 277-8581, Japan}

\author{Takahiro Misawa}
\affiliation{Institute for Solid State Physics, University of Tokyo, Kashiwa, 277-8581, Japan}

\begin{abstract}
Spin current, a key concept in spintronics that carries spin angular momentum, has a non-unique definition due to the non-conservation of spins in solids. While two primary definitions exist--- conventional spin current and conserved spin current---their validity has not been quantitatively examined. Here, we examine the validity of these definitions of spin current by comparing their spin Hall conductivities to the spin accumulation on edges of materials calculated in a real-time evolution simulation. Employing the Kane-Mele model with the Rashba term, which explicitly violates spin conservation, we reveal that the spin Hall conductivities calculated under both definitions fail to reproduce the simulated results of spin accumulation when the Rashba term is large. Our results suggest that the standard definitions of spin current and the associated spin Hall conductivity do not give an accurate quantitative estimate of spin accumulation. This conclusion indicates that real-time simulations are necessary to accurately estimate spin accumulation on edges/surfaces of materials.
\end{abstract}

\maketitle

{\it Introduction.---} Controlling and detecting spin degrees of freedom in materials has been a central issue in spintronics, aiming to develop next-generation devices that exploit spins \cite{wolf2001spintronics,bader2010spintronics,vzutic2004spintronics,fert2008nobel,hirohata2020review,dieny2020opportunities,bhatti2017spintronics}. Emerging from a fundamental understanding of spin dynamics and their manipulation, this field has potential applications in various areas such as data storage \cite{binasch1989enhanced,baibich1988giant,chappert2007emergence,puebla2020spintronic}, quantum computing \cite{awschalom2002semiconductor,burkard2000spintronics,he2019topological}, and sensors \cite{dieny1991giant,miyazaki1995giant,parkin2003magnetically,liu2019overview}. The concept of spin current, analogous to electronic current, plays a key role in understanding spin transport mechanisms \cite{ShiPRL2006,FuseyaJPSJ2016,maekawa2017spin,MurakamiScience2003,MurakamiPRB2004,SinovaPRL2004}. While the spin Hall effect (SHE) \cite{dyakonov1971current,hirsch1999spin,MurakamiScience2003,MurakamiPRB2004,SinovaPRL2004,kato2004observation,wunderlich2005experimental,sinova2015spin} and its inverse \cite{saitoh2006conversion,valenzuela2006direct,kimura2007room,seki2008giant} provide a method for detecting spin current, a proper definition of spin current remains a challenging issue due to the non-conservation of spins \cite{ShiPRL2006}. This difficulty arises from the spin-orbit interactions (SOI) inherent in solids, which disrupt spin conservation and make accurate characterization and manipulation of spin currents a complex task.

Conventionally, spin current (referred to as the conventional spin current) is defined as the product of the spin operator and the velocity operator, analogous to the definition of electric current, which is given by the product of the electric charge and the velocity operator \cite{MurakamiScience2003,MurakamiPRB2004,SinovaPRL2004}. However, this conventional definition assumes conservation of spin, even though the assumption does not hold in general. It has been pointed out that the definition leads to unphysical results, such as a finite spin current even in equilibrium states~\cite{RashbaPRB2003}. To resolve these issues, a concept of conserved spin current has been proposed, explicitly considering the conservation laws of spin in solids~\cite{ShiPRL2006}. Although it has been shown that the conserved spin current gives physically reasonable results in several systems, a thorough examination of its applicable range has not been performed. Specifically, it is not yet clear to what extent the spin Hall conductivity (SHC) calculated under conserved spin current quantitatively reproduces spin accumulation at edges or surfaces.

In this letter, we address this issue by focusing on spin accumulation in the Kane-Mele model~\cite{KaneMelePRL2005(1),KaneMelePRL2005(2)}, which is a canonical model of a quantum spin Hall insulator. In this model, the SHC and its corresponding spin accumulation are quantized when the Rashba term, which explicitly violates spin conservation, is absent. In this case, no matter which definition of spin current (conventional or conserved spin current) is used, the SHC exhibits the same quantized value. However, introducing the Rashba term may break this agreement. Here, to evaluate the accuracy of SHC induced by SHE, we directly solved the time-dependent Schr\"{o}dinger equations and estimated the spin accumulation at edges/surfaces. We then compared those values with the respective SHCs obtained from both definitions of spin current. As a result, we found that the conventional spin current does not give qualitatively correct results even when the Rashba term is small. In contrast, the conserved spin current gives quantitatively correct results when the Rashba term is small, but, as the term increases, the conserved spin current fails to reproduce the spin accumulation even at a qualitative level. These results suggest that both primary definitions of spin current fail to reproduce spin accumulation when the non-conservation of spins is significant.

\begin{figure*}[bt]
\begin{center}
\includegraphics[width=17.8cm]{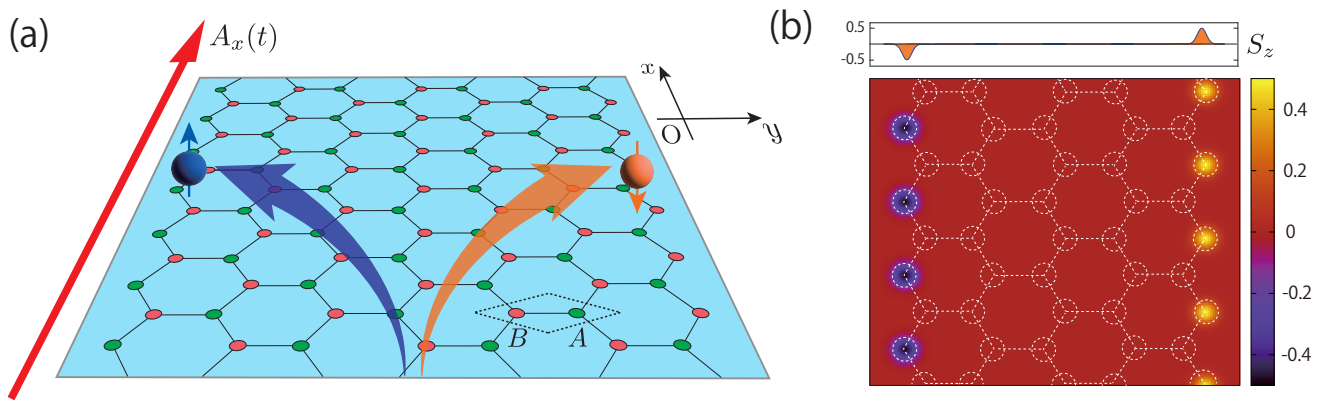}
\caption{(Color online) (a): Schematic illustration of SHE in the Kane-Mele model, where sublattice A (B) is indicated by the green (red) circles. The vector potential $A_{x}(t)$ is aligned along the $x$-axis, mirroring the zig-zag orientation of the system. (b): Numerical results displaying the real-space distribution of spin amplitude $\langle {s^{i_y}_z}\rangle$ on each atom for $t=T_{0}$. Here, we posit $\lambda_{so}=0.06t_{0}$, $\lambda_{v}=0$, and $\lambda_{R}=0$.
\label{fig:config1}}
\end{center}
\end{figure*}

{\it Model and method---} We start from the Kane-Mele model~\cite{KaneMelePRL2005(1),KaneMelePRL2005(2)}, a tight-binding model on a honeycomb lattice with SOI. Additionally, we introduce a staggered potential and the Rashba effect induced by a perpendicular static electric field. The total Hamiltonian of this model is given by
\begin{align}
H &= t_{0} \sum_{\langle i,j\rangle}c_{i}^\dagger c_{j} -i\lambda_{\rm SO}\sum_{\langle\langle i,j\rangle\rangle}\nu_{ij}c_{i}^{\dagger}s_{z} c_{j} \nonumber \\
&+ \lambda_{v} \sum_{i}\xi_i c_{i}^\dagger c_{i}
-i\lambda_{\rm R}\sum_{\langle i,j\rangle} c_{i}^{\dagger}(\bm{s} \times \hat{\bm{u}}_{ij}) c_{j} ,
\label{RS_Hamiltonian}
\end{align}
where $c_i$ ($c_i^\dagger$) is the annihilation (creation) operators of electrons, ${\bm{s}} = (s_x, s_y, s_z)$ is the spin operator, $\langle i,j\rangle$ and $\langle\langle i,j\rangle\rangle$ denotes a pair of nearest and next-nearest sites, $t_{0}$ is the hopping energy; the magnitudes of the staggered potential, SOI, and Rashba effect are respectively denoted as $\lambda_{v}$, $\lambda_{SO}$, and $\lambda_{R}$. The variable $\nu_{ij}$ takes $+1$($-1$) for a right (left) turn hopping from site $i$ to $j$ through an intermediate site, $\xi_i$ takes $+1$($-1$) on sublattice A (B), and $\hat{\bm{u}}_{ij}$ is a unit vector pointing from site $i$ to $j$.

To investigate SHE in the Kane-Mele model, we apply a DC electric field in the zig-zag direction ($x$-direction) to a system with a finite width in the armchair direction ($y$-direction). In this setup, spins accumulate around the zig-zag edge through the SHE (Fig.~\ref{fig:config1}(a)). Here, we impose a periodic boundary condition in the $x$-direction and introduce the DC electric field by using the Peierls phase~\cite{peierls1933theorie,peierls1955quantum}, i.e., replacing the $x$-component of the Bloch wavevector with $k'_{x}=k_{x}+ e A_{x}(t)$, where $A_{x}(t)$ is the vector potential. By taking a Fourier transformation, we obtain the Hamiltonian in a mixed representation, i.e., with the wavenumber in the $x$-direction, $k_x$, and the real-space site index in the $y$-direction, $i_y$ (the explicit form of the Hamiltonian is given in Supplementary Material Section I).

By solving the following time-dependent Schr\"{o}dinger equation, 
\begin{align}
i\frac{\partial \ket{\phi(t)}}{\partial t}=H(t)\ket{\phi(t)},
\end{align}
we can determine the time evolution of the wavefunction at each site $i_y$ for a specified $k_{x}$. Using $\ket{\phi(t)}$, we can calculate the time dependence of the spin polarization $\langle s^{i_y}_{z}\rangle$ at each site $i_y$. The total spin accumulation $\langle S^{\rm tot}_{z} \rangle$ is defined by summing these polarizations over half of the region in the $y$-direction. Throughout this work, we will set the magnitude of SOI to $\lambda_{SO}=0.06 t_{0}$ and the vector potential $A_x(t)$ to $2\pi t/L_{y}T_{0}$, where $L_{y}$ is the number of unit cells aligned along the y-direction. We will also set $\hbar=c=1$. The SHC $\sigma^{\rm{time}}_{s}(\lambda_{v},\lambda_{R})$ can be obtained from $\langle S^{\rm tot}_{z}\rangle$ at $t=T_{0}$ when a sufficiently large $T_0$ is chosen (for details, see Supplementary Material Section II-A).

The SHC for a bulk system can be calculated, using linear response theory, from the correlation function including
the spin current~\cite{Shitade2022}. 
The conventional and conserved spin currents, $j^{s}_y(t)$ and $J^{s}_y(t)$, have the following forms:~\cite{MurakamiPRB2004,MurakamiPRL2006,MurakamiScience2003,SinovaPRL2004,ShiPRL2006}
\begin{align*}
&j^{s}_y(t) = \frac{1}{2}(v_y s_z + s_z v_y), \\
&J^{s}_y(t) =\frac{1}{2}\frac{d}{dt}(y s_z + s_z y),
\end{align*}
where $v_y$ is the velocity operator that is defined by the time derivative of $y$. 
In the calculations of SHC, we employ the Hamiltonians of the Kane-Mele model in momentum space; i.e., we impose 
periodic boundary conditions in both the $x$- and $y$-directions.
Detailed expressions for SHC are given in Supplemental Material Section III-A and III-B.

\begin{figure}[t!]
\begin{center}
\includegraphics[width=8.2cm]{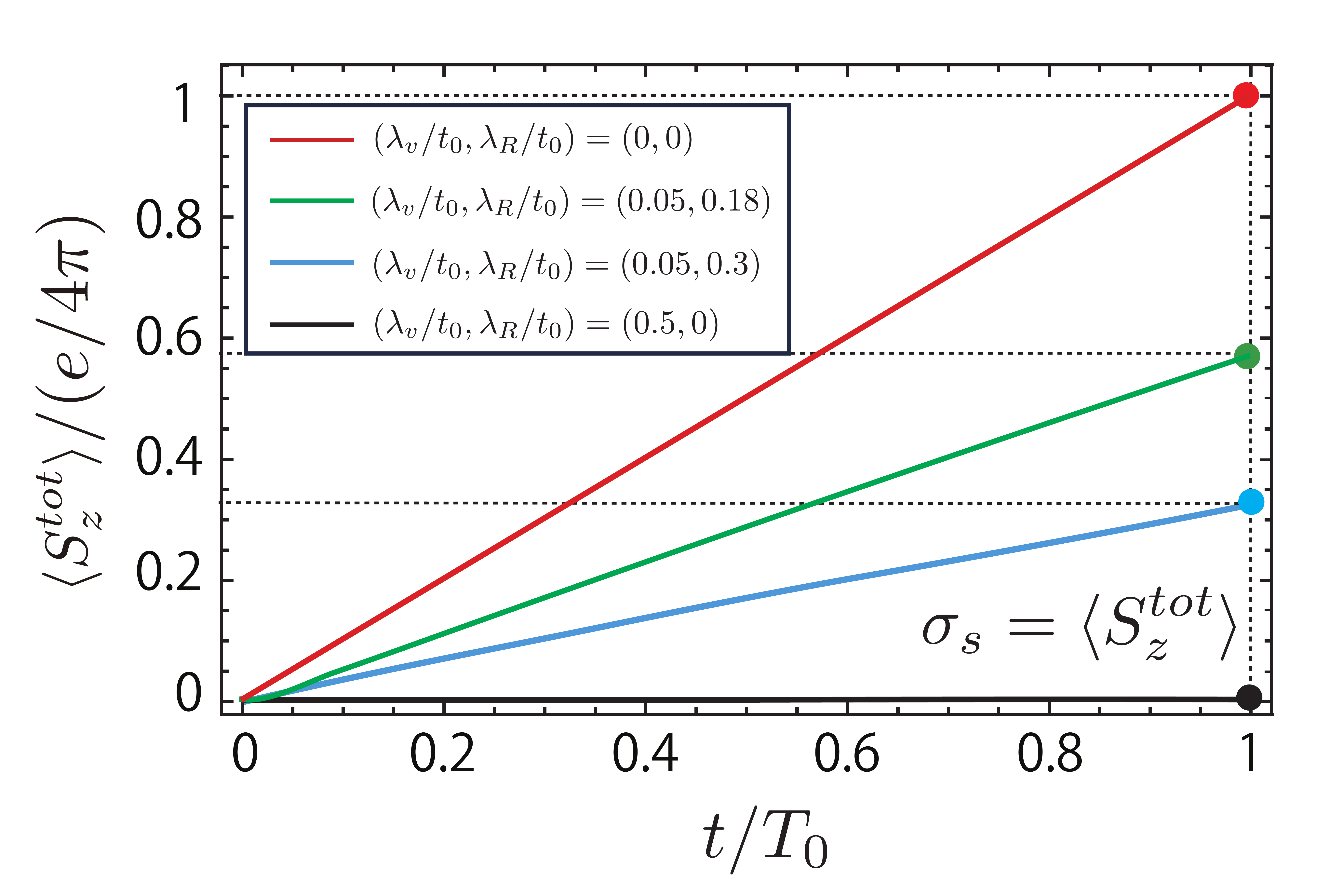}
\caption{(Color online) Numerical results of the time evolution of total spin accumulation on the edge of graphene, which is aligned along the $x$-axis. The red, green, blue, and black lines correspond to $(\lambda_{v}/t_0,\lambda_{R}/t_0)$ values of (0, 0), (0.05, 0.18), (0.05, 0.3), and (0.5, 0), respectively. The SHCs are estimated from $\left<S^{tot}_{z}\right>$ at $t=T_{0}$
\label{fig:config2}}
\end{center}
\end{figure}

\begin{figure*}[bt]
\begin{center}
\includegraphics[width=17.8cm]{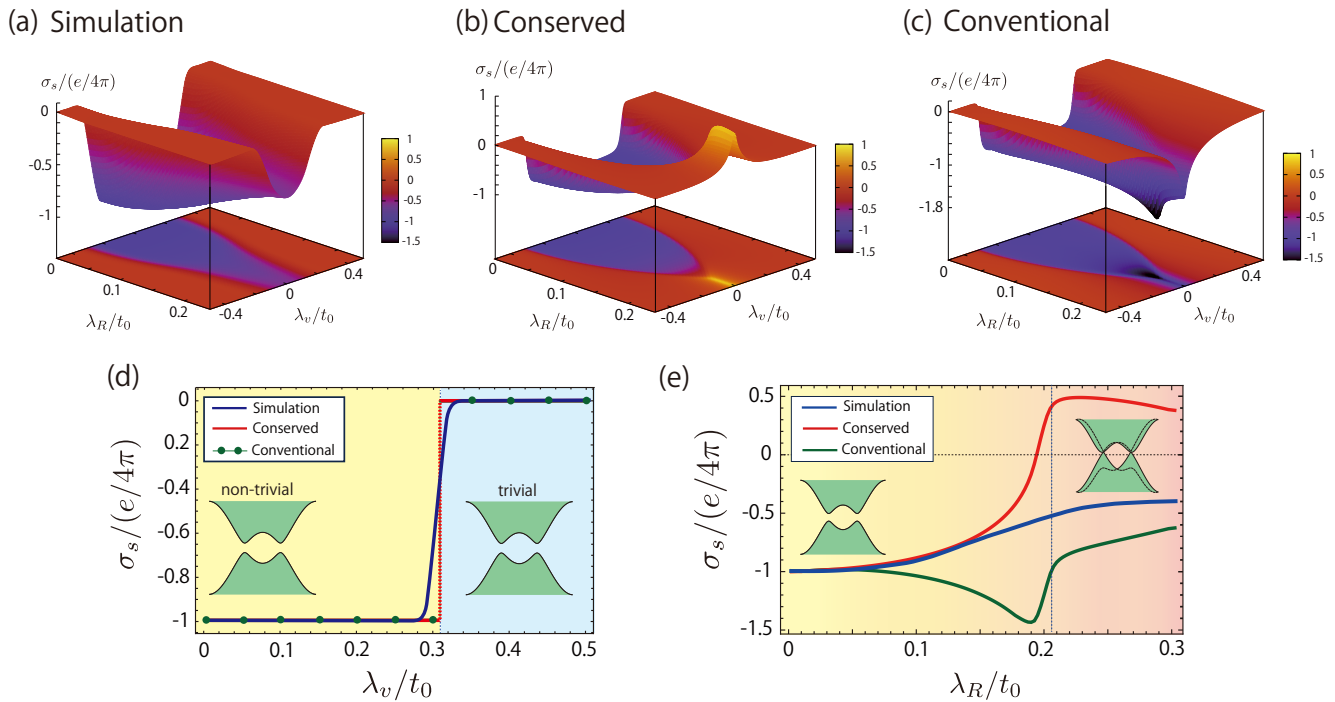}
\caption{(Color online) (a)-(c): Numerical results for SHC as a function of the Rashba effect $\lambda_{R}$ and staggered potential $\lambda_{v}$. Figure (a) illustrates results obtained from the simulation of the real-time evolution. Figures (b) and (c) represent calculations based on linear response theory and the definitions of conserved and conventional spin currents. (d) and (e): SHC plotted against a staggered potential $\lambda_{v}$ and the Rashba effect $\lambda_{R}$ for $\lambda_{R}/t_{0}=0$ and $\lambda_{v}/t_{0}=0.025$, respectively. The red and green lines are calculations from linear response theory using the definitions of conserved and conventional spin currents. The blue lines depict results from the simulated dynamics. 
\label{fig:config3}}
\end{center}
\end{figure*}

{\it Results---} In the Kane-Mele model without the Rashba term, the SOI opens a band gap and realizes a quantum spin Hall insulator in which SHC is quantized as $e/4\pi$. Figure~\ref{fig:config1}(b) shows how the spin accumulation $\langle s^{i}_z \rangle$ varies in real space at $t=T_0$ for $\lambda_{R}=0$ and $\lambda_{v}=0$. This figure indicates that spin accumulation predominantly occurs only on atoms at the edge. This result means that the total spin accumulation $\langle S^{\text{tot}}_{z} \rangle$ can be estimated by the spin of only one atom near the edges.

As shown in Fig.~\ref{fig:config2}, the spin accumulation around the edge in the quantum spin Hall insulator has a linear time dependence and is quantized (red line). As the staggered potential $\lambda_{v}$ increases, the band-gap energy decreases and becomes zero at a certain potential value, i.e., $\lambda_v=3\sqrt{3}\lambda_{SO}$ (see Supplementary Materials Section III-D). Above this threshold, the system enters a trivial phase in which the SHC vanishes. As shown by the black line in the figure, spin accumulation does not occur in a trivial insulator.

Variations in spin accumulation $\langle S^{\rm tot}_{z}\rangle$ induced by the Rashba term $\lambda_{R}$ are also shown in Fig.~\ref{fig:config2}. Notably, even in the presence of the Rashba term, the spin accumulation has a nearly linear time dependence, which indicates a linear response to the external DC electric field. The green and blue lines in the figure correspond to $\lambda_{R}/t_0=0.18$ and 0.3 when $\lambda_v$ is set to $0.05t_0$. We thus find that the Rashba term reduces spin accumulation at the edges. These results indicate that the SHC can be evaluated from the spin accumulation without ambiguity by using the real-time evolution.

Figure~\ref{fig:config3}(a) plots the SHC obtained from the real-time simulation ($\sigma_{s}^{\rm time}$) as a function of the Rashba term $\lambda_{R}$ and staggered potential $\lambda_{v}$. For comparison, Figs.~\ref{fig:config3}(b) and (c) display SHCs calculated from linear response theory based on conserved ($\sigma_{s}^{\rm cons}$) and conventional spin current ($\sigma_{s}^{\rm conv}$), respectively. Note that the SHC calculated using the conserved spin current (Fig.~\ref{fig:config3}(b)) is consistent with that of the previous study \cite{MurakamiPRL2006}. This shows that all SHCs are quantized in the quantum spin Hall insulator when the $z$-component of spin $s_{z}$ is conserved, i.e., $\lambda_{R}=0$. The quantum phase transition between the topological phase and the trivial phase at $\lambda_{R}=0$ is also captured at $\lambda_v=3\sqrt{3}\lambda_{SO}$ irrespective of the definition of SHC.

A significant difference arises around the quantum phase transition between the $Z_{2}$ topological insulator and the trivial insulator ($\lambda_{v}=0$ and $\lambda_{R}=2\sqrt{3}\lambda_{\rm SO}$). 
We find that $\sigma_{s}^{\rm cons}$ becomes positive around the phase transition point, while $\sigma_{s}^{\rm time}$ and $\sigma_{s}^{\rm conv}$ remain negative. We also find that $\sigma_{s}^{\rm conv}$ deviates from $\sigma_{s}^{\rm time}$ as $\lambda_{R}$ becomes larger.

To examine the above-mentioned difference quantitatively, we depict the SHC as a function of $\lambda_v/t_0$ [$\lambda_R/t_0$] in Fig.~\ref{fig:config3}(d) [(e)] for $\lambda_{R}/t_{0}=0$ [$\lambda_{v}/t_0=0.025$] \footnote{For these figures, we checked the system-size dependence of the real-time simulation and confirm their credibility (see Supplementary Material Section II-B).}. In Fig.~\ref{fig:config3}(d), $\sigma_{s}^{\rm time}$ coincides with $\sigma_{s}^{\rm conv}$ and $\sigma_{s}^{\rm cons}$ calculated from linear response theory. This confirms that both linear response theory and the real-time evolution give accurate predictions for SHC in the spin-conserved system.

Conversely, in the spin-non-conserved system, $\sigma_{s}^{\rm time}$ (the blue line in Fig.~\ref{fig:config3}(e)) generally differs from $\sigma_{s}^{\rm cons}$ (the red line) and $\sigma_{s}^{\rm conv}$(the green line). All SHCs agree with each other in the small $\lambda_{R}$ region, but they begin to deviate around $\lambda_{R}=0.1$. In particular, $\sigma_{s}^{\rm conv}$ exceeds the quantization value above $\lambda_{R}/t_0=0.1$. In contrast, $\sigma_{s}^{\rm cons}$ (the red line in Fig.~\ref{fig:config3}(e)) is consistent with $\sigma_{s}^{\rm time}$ up to around $\lambda_{R}/t_0=0.15$. Around the transition point, $\sigma_{s}^{\rm cons}$ reverses sign, whereas $\sigma_{s}^{\rm conv}$ exhibits a minimum value and maintains a negative value. These behaviors sharply contrast with that of $\sigma_{s}^{\rm time}$, which shows a continuous change as a function of $\lambda_{\rm R}$. These results indicate that the SHCs calculated under either conserved or conventional spin current do not quantitatively reproduce the simulated spin accumulation when the amplitude of the Rashba term is large, as in the broad parameter region $(\lambda_{v}, \lambda_{R})$ shown in Fig.~\ref{fig:config3}(a)-(c).

{\it Discussion---} The seminal paper by Fu and Mele~\cite{Fu_PRB2006} discussed spin accumulation in the Kane-Mele model by introducing the concept of a $Z_{2}$ spin pump. There, it was argued that the $Z_{2}$ spin pump could indeed induce spin accumulation at the edges of the system. In Ref.~\cite{Sheng_PRL2006}, spin accumulation was shown to occur even when the Rashba term exists and breaks spin conservation. On the other hand, the SHC, $\sigma_{s}$, formulated using linear response theory for a bulk system with conserved spin current, gives a finite value for the same system~\cite{MurakamiPRL2006}, which is consistent with the above discussions on spin accumulation. While the thus-formulated SHC $\sigma_{s}$ has often been used to analyze the spin accumulation of the spin-non-conserved systems \cite{MonacoPRB2020}, a quantitative comparison with actual spin accumulation has not been made so far. Our study clarifies that the SHC $\sigma_{s}$ obtained from linear response theory is not suited to quantitative evaluations of spin accumulation when the Rashba term becomes significant.

The previous studies on the Rashba model also pointed out that the spin accumulation can be finite~\cite{Shitade2022} even when the conserved and conventional spin currents yield zero SHC~\cite{Sugimoto_PRB2006}. We also should be careful of the fact that these calculations use biased approximations, such as the Born approximation. Our computational method enables accurate evaluations of spin accumulation in an unbiased way without depending on specific approximations. Our results also suggest that, regardless of which definition of spin current (conserved or conventional) is employed, it fails to accurately describe spin accumulation even in the simple Kane-Mele model with the Rashba term. This conclusion implies the necessity of utilizing real-time simulation calculations for making qualitative comparisons with experimental measurements of magnetization accumulation on material surfaces. An alternative way is to find a new definition of spin current that reproduces the spin accumulation obtained in real-time evolution. This is an intriguing issue that deserves further study.

{\it Conclusion---} We theoretically investigated the applicable range of the concepts of conventional and conserved spin currents. The spin current can be primarily estimated from the spin accumulated at the edge or surface. By employing the Kane-Mele model and conducting real-time evolution simulations, we concluded that the conventional spin current does not yield qualitatively correct results for small Rashba terms, while the conserved spin current provides qualitatively correct results. However, as the Rashba term increases, the conserved spin current also fails to reproduce the spin accumulation even at a qualitative level. Our numerical results suggest that real-space simulation calculations are necessary for estimating spin accumulation at the edges/surfaces and that one must be aware that the concept of spin current is merely a tool for assisting in physical understanding.

\begin{acknowledgments}
The authors express their gratitude to A. Shitade for fruitful discussions. T.T. acknowledges funding from JST PRESTO (JPMJPR2107).
\end{acknowledgments}

\bibliography{./reference}

\clearpage
\end{document}


\title{Supplementary Material for ``What is a proper definition of spin current? -- Lessons from the Kane-Mele Model''}

\author{Tomohiro Tamaya}
\email{tamaya@g.ecc.u-tokyo.ac.jp}
\affiliation{Institute for Solid State Physics, University of Tokyo, Kashiwa, 277-8581, Japan}
\affiliation{JST, PRESTO, 4-1-8 Honcho, Kawaguchi, Saitama, 332-0012, Japan}

\author{Takeo Kato}
\affiliation{Institute for Solid State Physics, University of Tokyo, Kashiwa, 277-8581, Japan}
\date{\today}

\author{Takahiro Misawa}
\affiliation{Institute for Solid State Physics, University of Tokyo, Kashiwa, 277-8581, Japan}

\maketitle

\section{Hamiltonian}

Here, we describe the Hamiltonian for the Kane-Mele model \cite{KaneMele1_2005,KaneMele2_2005} incorporating the staggered potential, spin-orbit interaction (SOI), and Rashba effect. First, we introduce the real-space representation of the Hamiltonian (Sec.~\ref{subsec1}). Next, we derive a mixed representation of the Hamiltonian by imposing a periodic boundary condition in the zig-zag direction and open boundary conditions in the armchair directions (Sec.~\ref{subsec2}). Finally, we explain how to treat the DC electric field applied along the zig-zag direction (Sec.~\ref{subsec3}).

\begin{figure}[b]
\centering
\includegraphics[width=16cm]{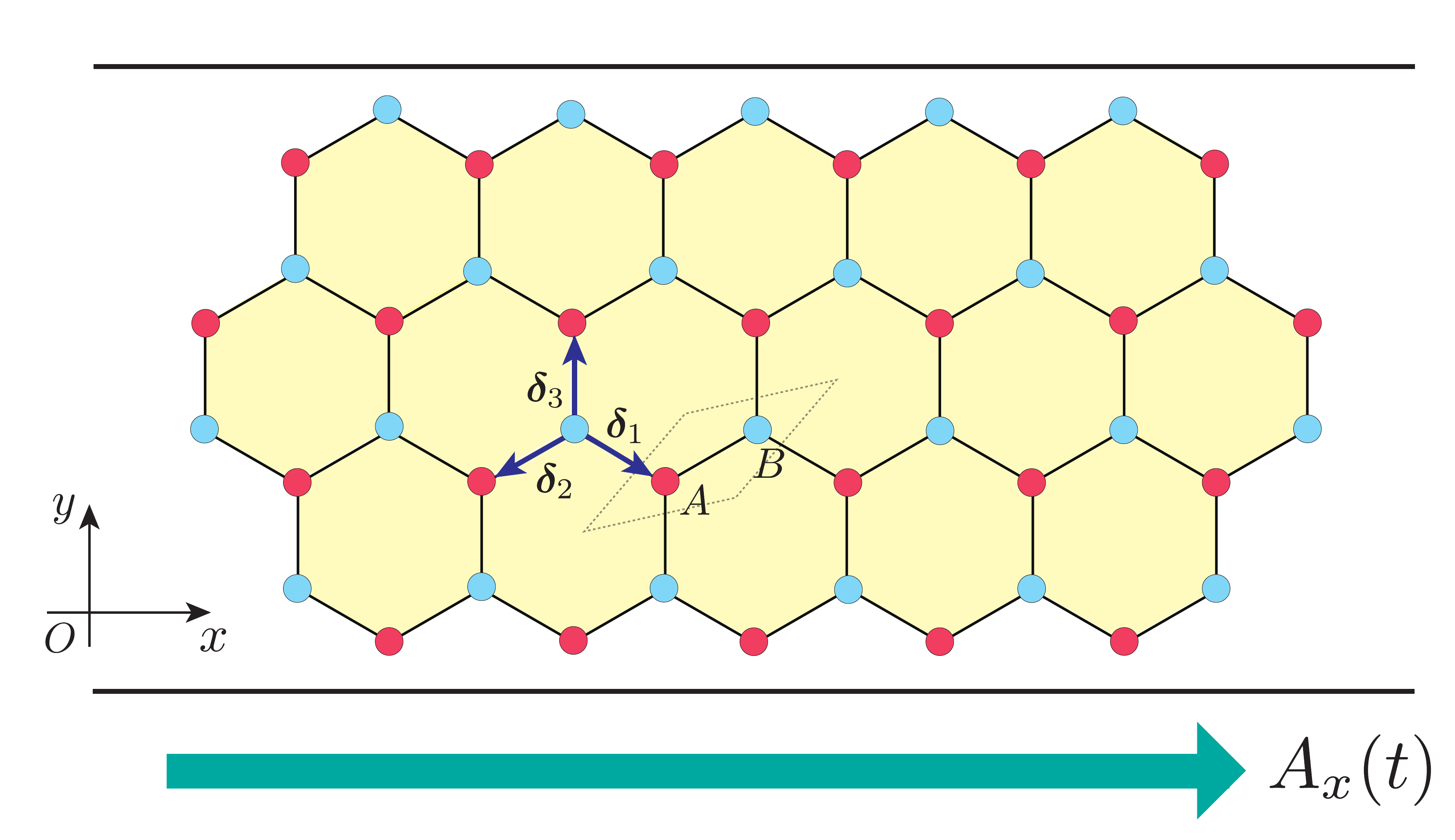}
\caption{(Color online) Schematic illustration of the lattice structure of the Kane-Mele model. In this configuration, periodic and open boundary conditions are imposed along the zig-zag (${\it{x}}$-axis) and armchair (${\it{y}}$-axis) directions, respectively. The three displacement vectors are $\bm{\delta}_{1} = a/2(\sqrt{3}, -1)$, $\bm{\delta}_{2} = a/2(-\sqrt{3}, -1)$, and $\bm{\delta}_{3} = a(0, 1)$, where $a$ is the lattice constant. An external DC electric field is applied along the ${\it{x}}$-axis.}
\label{figS1}
\end{figure}

\subsection{Real-space Representation}
\label{subsec1}
We consider a two-dimensional tight-binding model on a honeycomb lattice, as illustrated in Fig.~\ref{figS1}, where the unit cell comprises two sublattices labeled A and B, as depicted within the dotted rhomboid area. The total Hamiltonian is
\begin{align}
H &= H_{0} + H_{v} + H_{\rm SO} + H_{\rm R}, \label{RS_Hamiltonian}
 \\
H_{0} &= t_{0}\sum_{\langle i,j\rangle} (a_{i}^\dagger b_{j} + b_{j}^\dagger a_{i}) , \\
H_{v} &= \lambda_{v}\sum_{i}(a_{i}^\dagger a_{i} - b_{i}^\dagger b_{i}), \\
H_{\rm SO} &= -i\lambda_{\rm SO}\sum_{\langle\langle i,j\rangle\rangle}\nu_{ij}a_{i}^{\dagger}s_{z} a_{j} - i\lambda_{\rm SO}\sum_{\langle\langle i,j\rangle\rangle}\nu_{ij}b_{i}^{\dagger}s_{z} b_{j} , \\
H_{\rm R} &= -i\lambda_{\rm R}\sum_{\langle i,j\rangle}a_{i}^{\dagger}(\bm{s} \times \bm{u}_{ij}) b_{j} -i\lambda_{\rm R}\sum_{\langle i,j\rangle}b_{i}^{\dagger}(\bm{s} \times \bm{u}_{ij}) a_{j} , \label{HamR}
\end{align}
where $a_{i}=(a_{i\uparrow},a_{i\downarrow})$ and $b_{i}=(b_{i\uparrow},b_{i\downarrow})$ are the annihilation operators of electrons on sublattices A and B with spin-up and -down components, the indices for the sum $\langle i,j \rangle$ and $\langle \langle i,j \rangle \rangle$ indicate a pair of nearest- and next-nearest sites, $t_{0}$ is the hopping energy, and $\lambda_{v}$, $\lambda_{SO}$, and $\lambda_{R}$ are the magnitudes of the staggered potential, SOI, and Rashba effect, respectively. The vector $\bm{u}_{ij} = \bm{R}_{i}-\bm{R}_{j}$ denotes the displacement from site $i$ to $j$ and the variable $\nu_{ij}$ takes a value of $\pm 1$ depending on whether there is a right or a left turn hopping from site $i$ to $j$ via an intermediate site. The spin operators, ${\bm s}=(s_{x}, s_{y},s_{z})$, are expressed by the Pauli matrices:
\begin{align*}
s_{x}=\begin{pmatrix}0 & 1 \\ 1 & 0 \end{pmatrix}, \quad
s_{y}=\begin{pmatrix} 0 & -i \\ i & 0 \end{pmatrix}, \quad
s_{z}=\begin{pmatrix} 1 & 0 \\ 0 & -1 \end{pmatrix}.
\end{align*}
Note that the Rashba term $H_{\rm R}$ does not conserve the $z$-component of the total spin, while the other terms conserve it.

\subsection{Mixed Representation}
\label{subsec2}
As we are considering spin accumulation at the boundary in the $y$ (armchair) direction under a static electric field in the $x$ (zigzag) direction, we will impose periodic and open boundary conditions in the $x$- and $y$-directions, respectively. Applying Bloch's theorem in the $x$-direction, we can rewrite the Hamiltonian into a mixed representation, where the electronic states are assigned in accordance with the Bloch wavenumber in the $x$-direction, $k_x$, the real-space position of the unit cell in the $y$-direction, $j_y$, and the spin index, $\sigma$. Here, we introduce new creation operators of electrons,
\begin{align}
a^{\dagger}_{j\sigma} = \frac{1}{\sqrt{N}}\sum_{k_x}a^{\dagger}_{j_y \sigma k_x}e^{-i k_{x} R^{j_x}_{A}}, \qquad
b^{\dagger}_{j\sigma} = \frac{1}{\sqrt{N}}\sum_{k_x}b^{\dagger}_{j_y \sigma k_{x}}e^{-i k_{x} R^{j_x}_{B}}, 
\label{FT}
\end{align}
where $R^{j_x}_{\rm A}$ ($R^{j_x}_{\rm B}$) denotes the $x$-component of the position of the site on sublattices A (B) for the $j$-th unit cell. In the subsequent subsections, the four terms in the Hamiltonian (\ref{RS_Hamiltonian}) are rewritten with this mixed representation. In the following, we use displacement vectors spanning neighboring sites, which are defined as ${\bm{\delta}_{1}}=a/2(\sqrt{3},-1)$, ${\bm{\delta}}_{2}=a/2(-\sqrt{3},-1)$, and ${\bm{\delta}_{3}}=a(0,1)$, with $a$ being the lattice constant (see Fig.~\ref{figS1}).

\subsubsection{Hopping and staggered potential terms}
The hopping term $H_0$ is rewritten as
\begin{align}
H_{0} &= t_{0}\sum_{j_y,\sigma,k_x} \sum_{l=1}^3 [e^{-ik_{x}\delta^{x}_{l}}a_{j_y\sigma k_{x}}^{\dagger} b_{j_{yl}\sigma k_{x}} + {\rm h.c.} ],
\end{align}
where $\delta^{x}_{l}$ ($l=1,2,3$) is the $x$-component of the displacement vector ${\bm \delta}_l$ and $j_{yl}$ denotes the $y$-component of the position of the unit cell when we start at site $j$ and move to the neighboring site specified by ${\bm \delta}_l$ ($l=1,2,3$). The staggered potential term $H_{v}$ can be rewritten as
\begin{align}
H_{v} = \lambda_{v}\sum_{j_y,\sigma,k_{x}}(a_{j_y \sigma k_{x}}^{\dagger} a_{j_y \sigma k_{x}} - b_{j_y \sigma k_{x}}^{\dagger} b_{j_y \sigma k_{x}}).
\end{align}

\subsubsection{Spin-orbit interaction term}
\label{subsubsec_HSO}

\begin{figure}[tb]
\centering
\includegraphics[width=6cm]{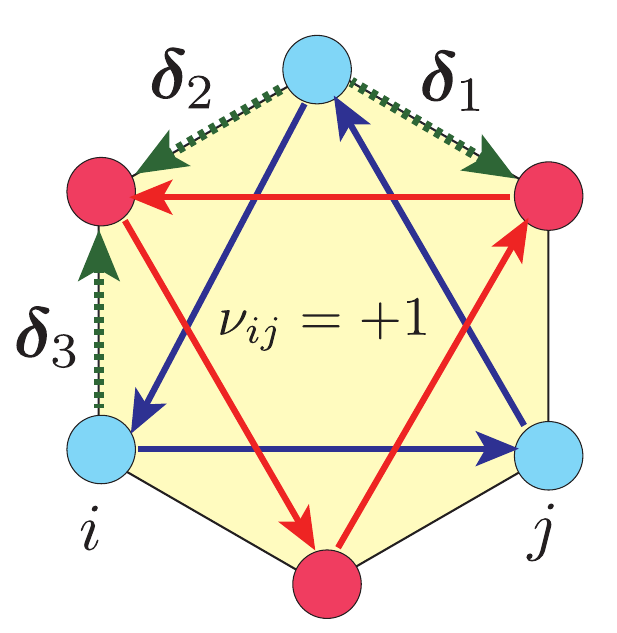}
\caption{(Color online) Schematic figure illustrating the hopping process in the SOI Hamiltonian. The red and blue circles represent sublattices A and B, respectively. The variable $\nu_{ij}$ takes a value of $\pm 1$ depending on whether there is a right or a left turn hopping from site $i$ to $j$ via an intermediate site.}
\label{figS2}
\end{figure}

By recovering the spin index, the Hamiltonian for the SOI, $H_{\rm SO}$, can be expressed in the real-space representation as
\begin{align}
H_{\rm SO} &= -i\lambda_{\rm SO}\sum_{\langle\langle i,j\rangle\rangle, \sigma}\nu_{ij}\sigma a_{i\sigma}^{\dagger} a_{j\sigma} - i\lambda_{\rm SO}\sum_{\langle\langle i,j\rangle\rangle,\sigma}\nu_{ij}\sigma b_{i\sigma}^{\dagger} b_{j\sigma},
\end{align}
where we have set $\sigma=+1$ ($-1$) for the spin-$\uparrow$ (spin-$\downarrow$) state. Note that the variable $\nu_{ij}$ always takes $+1$ for the hop path shown in Fig.~\ref{figS2}. The Hamiltonian $H_{\rm SO}$ can be rewritten in the mixed representation as
\begin{align*}
H_{\rm SO} &= -i \lambda_{\rm SO}\sum_{j_y,\sigma,k_x}\sum_{l'=1}^3 F(k_x)\sigma b^{\dagger}_{j_{y} \sigma k_x} b_{j_{l'y} \sigma k_x} -i \lambda_{\rm SO}\sum_{j_y, j_y',\sigma,k_x}\sum_{l'=1}^3 F^{*}(k_x)\sigma a^{\dagger}_{j_y \sigma k_x} a_{j_{l'y} \sigma k_x}, \\
F(k_x) &= -2 i [\sin (k_{x} (\delta^{x}_{2}-\delta^{x}_{3})) + \sin (k_{x} (\delta^{x}_{3}-\delta^{x}_{1})) + \sin(k_{x} (\delta^{x}_{1}-\delta^{x}_{2})) ] ,
\end{align*}
where $j_{l'y}$ denotes the position of the unit cell in the $y$-direction after moving from site $j$ to the next-nearest site through a displacement ${\bm \delta}_{l'}-{\bm \delta}_{l'+1}$ ($l'=1,2,3$), with ${\bm \delta}_{4}$ being ${\bm \delta}_{1}$.

\subsubsection{Rashba term}
By recovering the spin index, the first term in the Rashba Hamiltonian $H_{\rm R}$ given in Eq.~(\ref{HamR}) can be written as
\begin{align}
&-i\lambda_{\rm R}\sum_{\langle i,j\rangle}a_{i}^{\dagger}(\bm{s} \times \bm{u}_{ij}) b_{j} \nonumber \\
&= -i\lambda_{\rm R}\sum_{i} \sum_{l=1}^3
\delta^{y}_l(a_{i+\delta_l\uparrow}^{\dagger}b_{i\downarrow}+a_{i+\delta_l\downarrow}^{\dagger}b_{i\uparrow}) 
+\lambda_{\rm R}\sum_{i} \sum_{l=1}^3 \delta^{x}_l(a_{i+\delta_l\uparrow}^{\dagger}b_{i\downarrow}-a_{i+\delta_l\downarrow}^{\dagger}b_{i\uparrow}) .
\end{align}
This term can be rewritten in the mixed representation as
\begin{align}
&-i\lambda_{\rm R}\sum_{\langle i,j\rangle}a_{i}^{\dagger}(\bm{s} \times \bm{u}_{ij}) b_{j} \nonumber \\
&= \sum_{k_{x},j_y} \sum_{l=1}^3 \left[
-i\lambda_{\rm R}[\delta^{y}_{l}+i\delta^{x}_{l}]e^{-i k_{x} \delta^{x}_{l}}a_{j_{y}\uparrow k_{x}}^{\dagger} b_{j_{yl} \downarrow k_{x}} -i\lambda_{\rm R}[\delta^{y}_{l}-i\delta^{x}_{l}]e^{-i k_{x} \delta^{x}_{l}}a_{j_{y} \downarrow k_{x}}^{\dagger}b_{j_{yl} \uparrow k_{x}}\right] ,
\label{HamR1}
\end{align}
where $j_{yl}$ denotes the $y$-component of the position of the unit cell when we start at site $j$ and move to the neighboring site specified by ${\bm \delta}_l$ ($l=1,2,3$). Since the second term in the Rashba Hamiltonian (\ref{HamR}) is given by the Hermitian conjugate of Eq.~(\ref{HamR1}), the Rashba Hamiltonian $H_{\rm R}$ finally becomes
\begin{align}
H_{\rm R} &= \sum_{j_y,k_{x}} \sum_{l=1}^3 \left[ A(k_{x})a_{j_{y} \uparrow k_{x}}^{\dagger}b_{j_{yl} \downarrow k_{x}} + B(k_{x})a_{j_{y} \downarrow k_{x}}^{\dagger}b_{j_{yl} \uparrow k_{x}} + {\rm h.c.}\right], \\
A(k_{x}) &= -i\lambda_{\rm R} [\delta^{y}_{l}+i\delta^{x}_{l}]e^{-i k_{x} \delta^{x}_{l}}, \\
B(k_{x}) &= -i\lambda_{\rm R} [\delta^{y}_{l}-i\delta^{x}_{l}]e^{-i k_{x} \delta^{x}_{l}}.
\end{align}

\subsection{Application of DC Electric Fields}
\label{subsec3}
Next, we explain how to incorporate the effect of a DC electric field into the Hamiltonian. An adiabatic electric field can be introduced through the Peierls phase, where the Bloch wavevector $k_{x}$ is replaced by $k^{A(t)}_{x} \equiv k_{x} + eA_{x}(t)$. In our study, we perform this replacement on the single-particle and SOI Hamiltonians:
\begin{align}
H_{0} &= t_{0}\sum_{j_y,\sigma,k_{x}} {\sum_{l=1}^3} \left[f(k^{A(t)}_{x})a_{{j_{y}}\sigma k_{x}}^{\dagger} b_{{j_{ly}}\sigma k_{x}} + \text{h.c.}\right], \\
H_{\rm SO} &= -i \lambda_{\rm SO}\sum_{j_y,\sigma,k_x}
\sum_{l'=1}^3 F(k^{A(t)}_x) \sigma b^{\dagger}_{{j_{y}} \sigma k_x} b_{{j_{l'y}} \sigma k_x} -i \lambda_{\rm SO}\sum_{{j_y,\sigma,k_x}} {\sum_{l'=1}^3}F^{*}(k^{A(t)}_x)\sigma a^{\dagger}_{j_{ly} \sigma k_x} a_{j_{y} \sigma k_x}.
\end{align}
We have not used the Peierls phase on the Rashba Hamiltonian because spin-flip hopping cannot be induced by a DC driving force. The following section describes the computational methods using this Hamiltonian.

\section{Numerical Methods}

\begin{figure}[tb]
\centering
\includegraphics[width=16cm]{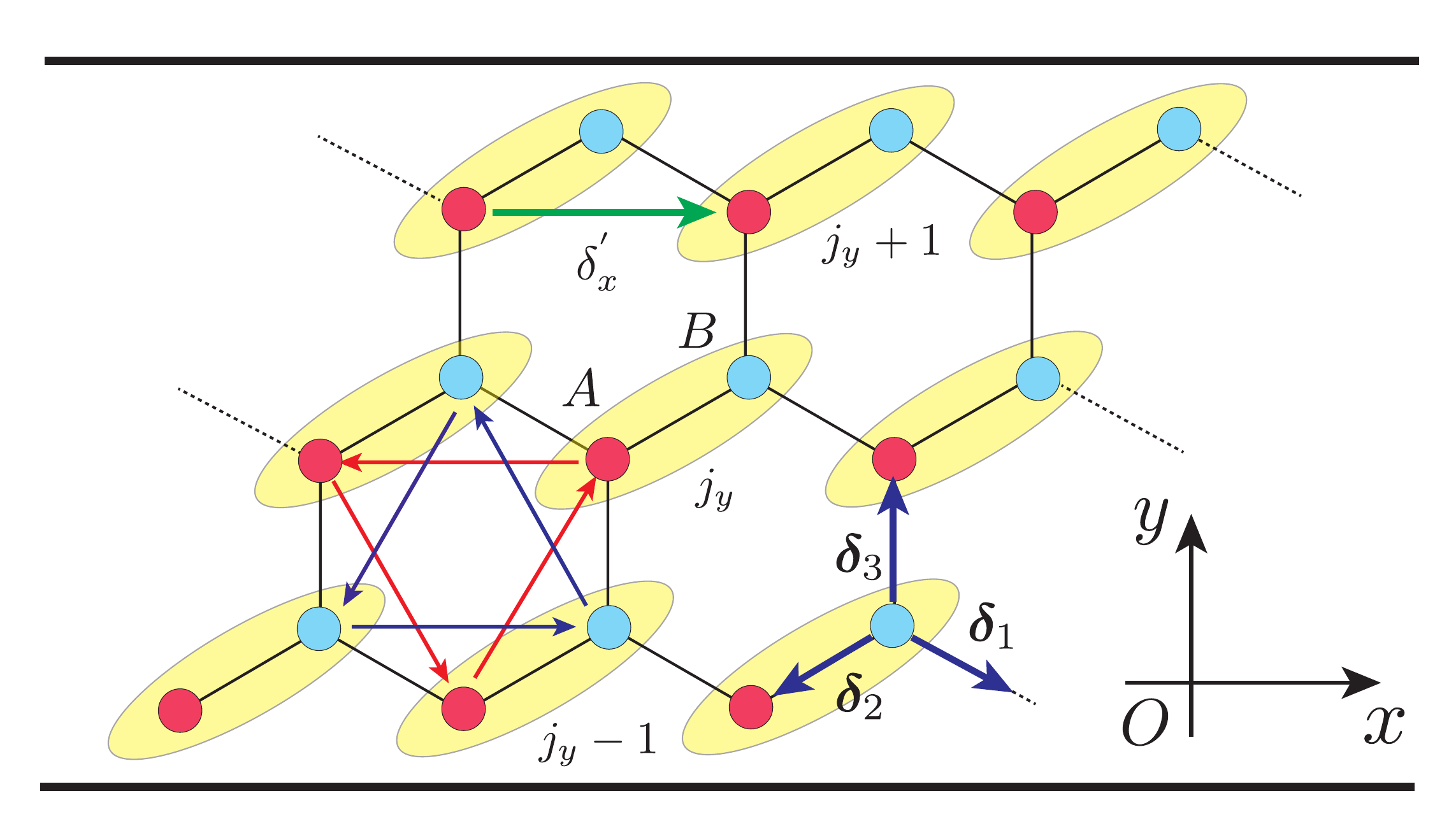}
\caption{(Color online) Schematic illustration of a finite-width honeycomb lattice structure. The yellow shaded area indicates the unit cell, with sublattices A and B represented by red and blue circles, respectively. Red and blue lines illustrate SOI hopping, while the green line denotes the unit cell's translation vector. The index $j_y$ indicates the position of the unit cell in the $y$-direction.}
\label{figS3}
\end{figure}

The unit cell in the Kane-Mele model is shown as the yellow area in Fig.~\ref{figS3}. We consider a finite-size lattice including $L_{x}$ and $L_{y}$ unit cells in the $x$- and $y$-directions. As a result of the periodic boundary conditions in the $x$-direction, $k_{x}$ is given by
\begin{align*}
k_{x}=\frac{2\pi n}{\delta'_{x} L_{x}}, \quad (n=0,1,2,\cdots, L_{x}-1),
\end{align*}
where $\delta'_{x}=|\delta^{x}_{1}-\delta^{x}_{2}|$ is the basic lattice vector pointing in the $x$-direction. For a fixed $k_x$, we only need to solve a one-dimensional lattice in the $y$-direction. Since there are four states in a unit cell, the corresponding Hamiltonian is represented by a $4L_{y} \times 4 L_{y}$ matrix.

\subsection{Numerical Calculation of Spin-Hall Conductivity}

We numerically solved the time-dependent Schr\"{o}dinger equation,
\begin{align*}
i \hbar \frac{d\ket{\psi}}{dt}=H\ket{\psi},
\end{align*}
by using the mixed representation in which the state $\ket{\psi}$ and the Hamiltonian $H$ are expressed as a $4L_y$-dimensional complex vector and a $4L_y\times 4L_y$ Hermitian matrix, respectively. By discretizing the time variable, we can numerically obtain time evolution from the initial state (the ground state) by using linear algebra. Then, we can calculate the averaged spin polarization on each unit cell $\langle s^{j_y}_z (t)\rangle$ at time $t$ and obtain the spin polarization, defined as 
\begin{align}
\langle S^{\rm edge}_z(t) \rangle = \sum_{j_y<L_{y}/2} \langle s^{j_y}_z(t) \rangle - \sum_{j_y>L_{y}/2} \langle s^{j_y}_z(t) \rangle
= 2 \sum_{j_y<L_{y}/2} \langle s^{j_y}_z (t)\rangle ,
\end{align}
where, on the second line, we have used the symmetry relation $\langle s^{j_y}_z(t) \rangle=-\langle s^{L_y-j_y}_z (t)\rangle$. Note that, in the actual simulation, the absolute value of $\langle s^{j_y}_z \rangle$ becomes large only near the edge (see Fig.~1(b) in the main text).

The procedure for obtaining the spin-Hall conductivity (SHC) $\sigma_{s}(\lambda_{v},\lambda_{R})$ from the real-time simulation is as follows. We apply a DC electric field by considering the time-dependent vector potential,
\begin{align}
A_{x}(t)=A_{0} t = \frac{2\pi t}{L_y T_0}.
\end{align}
The electric field is given as 
\begin{align}
 E_{0}=-\frac{dA(t)}{dt}=-\frac{2\pi}{L_y T_0}.
\end{align}
In the linear response regime, $\langle S^{\rm edge}_z \rangle=2\sum_{i_y<L_{y}/2}\langle s^{i_y}_z \rangle$ can be rewritten as
\begin{align}
\langle S^{\rm edge}_z(T_0) \rangle &= 2L_y \int_{0}^{T_0}dt \, j_{s}(t) = 2L_y\int_{0}^{T_0}dt \, \sigma_{s}(\lambda_{v},\lambda_{R}) E_{0} = 4\pi \sigma_{s}(\lambda_{v},\lambda_{R}),
\end{align}
where we have used $\langle S^{\rm edge}_z(0) \rangle=0$ for the initial ground state. Consequently, the SHC can be obtained by performing a real-time simulation of $\langle S^{\rm edge}_z(t) \rangle$ up to $t=T_{0}$.

\begin{figure}[tb]
\centering
\includegraphics[width=16.5cm]{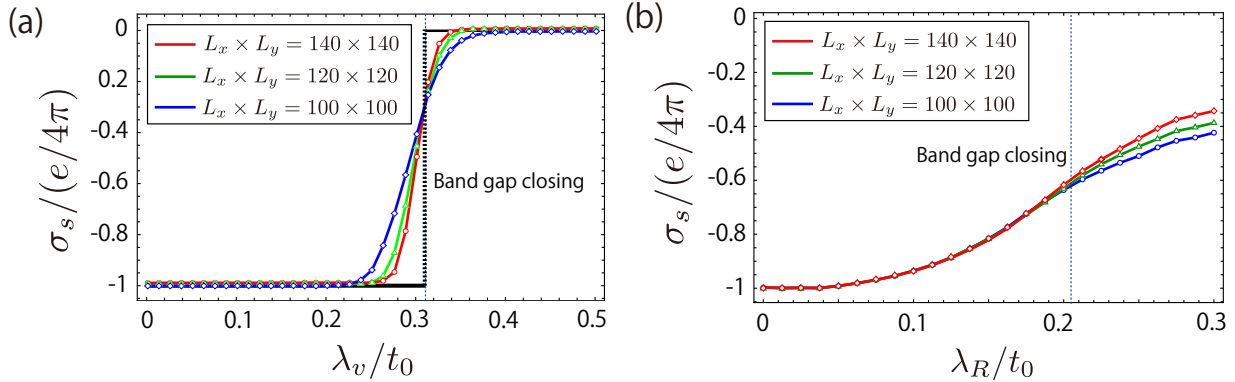}
\caption{(Color online) Numerical results of SHC as a function of (a) $\lambda_{v}$ at $\lambda_{\rm R}/t_{0}=0$ and (b) $\lambda_{\rm R}$ at $\lambda_{v}/t_{0}=0.025$ for different system sizes. The red, green, and blue lines indicate SHC for $L_{x}\times L_{y} =100 \times 100$, $120 \times 120$, and $140 \times 140$.}
\label{figS4}
\end{figure}

\subsection{System-Size Dependence of the Numerical Results}
Figure~\ref{figS4}(a) illustrates the numerical results for SHC as a function of $\lambda_{v}$ at $\lambda_{\rm SO}/t_{0}=0.06$ and $\lambda_{\rm R}/t_{0}=0$ for different system sizes. The red, green, and blue lines represent the SHC for $L_{x}\times L_{y} =100 \times 100$, $120 \times 120$, and $140 \times 140$, respectively. In this case, the $z$-component of the total spin is conserved. This figure indicates that the numerical results approach the value expected from linear response theory (depicted by the black line) as the system size increases. Figure~\ref{figS4}(b) illustrates numerical results for SHC as a function of $\lambda_{\rm R}$ at $\lambda_{v}/t_{0}=0.025$ for different system sizes. The red, green, and blue lines represent the result for $L_{x}\times L_{y} =100 \times 100$, $120 \times 120$, and $140 \times 140$, respectively. This figure also indicates that, as the system size becomes larger, the more evident the convergence is towards a certain value. In particular, the numerical results almost converge before the gap closes at $\lambda_{\rm R}/t_0 = 2\sqrt{3}\lambda_{\rm SO}$. Thus, we conclude that Fig. 3(d) and (e) in the main text provide credible numerical results for our discussion.

\section{Linear Response Theory for Conventional and Conserved Spin Current}
Here, we provide details on the formulation of SHC using linear response theory. We consider an external DC electric field applied in the $x$-direction, resulting in spin accumulation at the edge. This configuration anticipates the generation of a spin current along the $y$-axis. In Sec.~\ref{subsec2_1}, we introduce the definition of conventional spin current \cite{Murakami2003,Sinova2004,MurakamiPRB2004}. By employing linear response theory, we derive an expression for the SHC that corresponds to the Thouless-Kohmoto-Nightingale-Nijs (TKNN) formula \cite{TKNN1982} when spin is conserved. In Sec.~\ref{subsec2_2}, we define conserved spin current \cite{Shi2006} and show the corresponding formula for SHC. In Sec.~\ref{subsec2_3}, we construct a $\bm{k}$-space Hamiltonian for the Kane-Mele model and outline the procedure for obtaining Bloch wavefunctions and eigenvalues for the numerical calculation of SHC using conventional or conserved spin current. Finally, in Sec.~\ref{subsec2_4}, we refer to some properties of the band structures of the Kane-Mele model depending on the magnitude of the SOI, staggered potential, and Rashba effect.

\subsection{Conventional Spin Current}
\label{subsec2_1}
The concept of spin current is introduced based on the analogy to the concept of electric current; we assume that the spin current induces spin accumulation near the edge \cite{Murakami2003,Sinova2004,MurakamiPRB2004}. This naive consideration leads to the conventional definition of spin current $j^{s}_{y}(t)$:
\begin{align}
j^{s}_y(t) =\frac{\hbar}{2}(v_y s_z + s_z v_y),
\end{align}
where $v_{y}$ is the velocity operator defined by $v_{y}=\dot{y}=i\hbar [y,H]$. This definition is analogous to the electric current $j_{y}=e v_{y}$, where the electric charge $e$ is replaced by the spin operator $s_z$, taking into account the Hermitian conjugate, which requires an anti-commutation relationship of operators and a prefactor of $1/2$. By employing linear response theory based on this definition, we can obtain the SHC as \cite{Shitade2022}
\begin{align}
\sigma^{xy}_{s} &= -\frac{e}{\hbar}\sum_{n}\int \frac{d^2 k}{(2\pi)^2}\, b^{xy}_{n{\bm{k}}}f(\epsilon^{\bm{k}}_{n}),
\label{Conventional_1}\\
b^{xy}_{n{\bm{k}}} &= i \sum_{m(\neq n)}\frac{\braket{u^{\bm{k}}_n|j^{s}_y|u^{\bm{k}}_m}\braket{u^{\bm{k}}_m|\hbar v_{x}|u^{\bm{k}}_n}-{\rm c.c.}}{(\epsilon^{\bm{k}}_n-\epsilon^{\bm{k}}_m)^2},
\label{Conventional_2}
\end{align}
where $\epsilon^{\bm{k}}_{n}$ and $\ket{u^{\bm{k}}_n}$ are respectively the band energy and the Bloch wave function with Bloch wavenumber $\bm{k}$ and band index $n$, $\hbar$ is the reduced Planck constant, and $f(\epsilon)$ is the Fermi distribution function. In the present notation, we include the spin index $\sigma$ in the band index $n$.

Let us suppose a special case in which the Bloch wavefunctions are always eigenfunctions of the spin operators $s_{z}$, that is $s_z\ket{u^{\bm{k}}_{n\sigma}} = \pm \ket{u^{\bm{k}}_{n\sigma}}$ for spin-up and -down components. By separating the spin index $\sigma$ from the band index $n$, Eq.~(\ref{Conventional_2}) can be rewritten as
\begin{align}
b^{xy}_{n\bm{k}} &= -2\hbar \sum_{m(\neq n)}\left[\frac{{\rm{Im}}[\braket{u^{\bm{k}}_{n\uparrow}|j^{s}_y|u^{\bm{k}}_{m\uparrow}}\braket{u^{\bm{k}}_{m\uparrow}|v_{x}|u^{\bm{k}}_{n\uparrow}}]}{(\epsilon^{\bm{k}}_n - \epsilon^{\bm{k}}_m)^2} +\frac{{\rm{Im}}[\braket{u^{\bm{k}}_{n\downarrow}|j^{s}_y|u^{\bm{k}}_{m\downarrow}}\braket{u^{\bm{k}}_{m\downarrow}|v_{x}|u^{\bm{k}}_{n\downarrow}}]}{(\epsilon^{\bm{k}}_n-\epsilon^{\bm{k}}_m)^2}\right] \nonumber \\
& = -2\hbar \sum_{m(\neq n)}\left[\frac{{\rm{Im}}[\braket{u^{\bm{k}}_{n\uparrow}|v_y|u^{\bm{k}}_{m\uparrow}}\braket{u^{\bm{k}}_{m\uparrow}|v_{x}|u^{\bm{k}}_{n\uparrow}}]}{(\epsilon^{\bm{k}}_n-\epsilon^{\bm{k}}_m)^2} -\frac{{\rm{Im}}[\braket{u^{\bm{k}}_{n\downarrow}|v_y|u^{\bm{k}}_{m\downarrow}}\braket{u^{\bm{k}}_{m\downarrow}|v_{x}|u^{\bm{k}}_{n\downarrow}}]}{(\epsilon^{\bm{k}}_n-\epsilon^{\bm{k}}_m)^2}\right]. 
\label{Conventional_3}
\end{align}
Here, we use the following famous relations:
\begin{align}
&\braket{u^{\bm{k}}_{n\sigma}|v_i|u^{\bm{k}}_{m\sigma}}=\frac{(\epsilon^{\bm{k}}_n-\epsilon^{\bm{k}}_m)}{\hbar}\braket{u^{\bm{k}}_{n\sigma}|\frac{d}{d k_i}|u^{\bm{k}}_{m\sigma}}, \nonumber \\
&\Braket{\frac{d u^{\bm{k}}_{n\sigma}}{d k_{i}}|u^{\bm{k}}_{m\sigma}}=-\Braket{u^{\bm{k}}_{n\sigma}|\frac{d u^{\bm{k}}_{m\sigma}}{d k_{i}}},
\end{align}
where $i=x,y$. Substituting them into Eq. (\ref{Conventional_3}), we obtain 
\begin{align}
&\sigma^{xy}_{s} = \sum_{n} \int \frac{d^2 k}{(2\pi)^2}(\Omega_{n\uparrow}(\bm{k})-\Omega_{n\downarrow}(\bm{k}))f(\epsilon^{\bm{k}}_{n}), \nonumber \\
&\Omega_{n\sigma}(\bm{k}) = \frac{d}{dk_{x}}\Braket{u^{\bm{k}}_{n\sigma}|\frac{d}{d k_y}u^{\bm{k}}_{m\sigma}} - \frac{d}{dk_{y}}\Braket{u^{\bm{k}}_{n\sigma}|\frac{d}{d k_x}u^{\bm{k}}_{m\sigma}}.
\end{align}
By introducing the Berry connection and the Berry curvature,
\begin{align}
\bm{A}_{n\sigma}(\bm{k})& =\braket{u^{\bm{k}}_{n\sigma}|i\nabla_{{\bm k}}|u^{\bm{k}}_{n\sigma}}, \\
\bm{B}_{n\sigma}(\bm{k}) &=\nabla_{\bm k}\times \bm{A}_{n\sigma}(\bm{k}), 
\end{align}
respectively, the SHC can finally be expressed in the form,
\begin{align}
\sigma^{xy}_{s} = -\frac{e}{\hbar}\sum_{n}\int \frac{d^2 k}{(2\pi)^2}[\bm{B}_{n\uparrow}(\bm{k})-\bm{B}_{n\downarrow}(\bm{k})]_z f(\epsilon^{\bm{k}}_{n}) .
\label{Conventional_4}
\end{align}
Thus, the formula for the SHC is regarded as an expansion of the TKNN formula \cite{TKNN1982} for $\lambda_{R}=0$, where the phenomenological definition of the spin current, $j^{s}_{y}(t) \propto j_{y\uparrow}(t) - j_{y\downarrow}(t)$, with the electric current $j_{y\sigma}(t)$ carried by electrons with spin $\sigma$ is justified. We should note that this definition of spin current is frequently used in spintronics research in combination with the Boltzmann equation or spin diffusion theory. To obtain the numerical calculation depicted in Fig.~3(c) in the main text, we need to calculate $\sigma^{xy}_{s}$ of the SHC by reverting to Eq.~(\ref{Conventional_1}) because the finite Rashba effect breaks spin conservation.

\subsection{Conserved Spin Current}
\label{subsec2_2}
Several issues have been raised concerning the conventional spin current \cite{Shi2006}. Firstly, the definition relies on an analogy to electric current, derived from charge conservation. However, when spin $s_{z}$ is not conserved in materials, i.e., $[s_{z},H] \neq 0$, the analogy breaks down. Secondly, as highlighted by E. I. Rashba \cite{Rashba2003}, the conventional definition allows for a finite spin current to flow in Rashba systems even without applying a voltage. This observation raises concerns about the appropriateness of the spin current definition for describing real transport phenomena. Thirdly, the use of the conventional definition does not allow for the derivation of mechanical or thermodynamic forces associated with this flow. This suggests that describing spin transport using the conventional spin current definition does not align with near-equilibrium transport theory.

To overcome these shortcomings, a new definition of the spin current has been proposed \cite{Shi2006}. Here, the spin current flowing in the $y$-direction with a $s_z$ component is expressed as
\begin{align}
J^{s}_y(t) =\frac{1}{2}\frac{d}{dt}(y s_z + s_z y) =\frac{1}{2}(v_y s_z + s_z v_y + y \dot{s}_z +\dot{s}_z y) \equiv j^{y}_s(t) + P^{y}_{\tau}(t).
\label{SC_Conserved}
\end{align}
In this definition, called `conserved spin current', a new term $P^{y}_{\tau}$ appears in addition to those of the conventional definition of spin current $j^{y}_{s}(t)$. Using the relation,
\begin{align}
\frac{\partial P^{y}_{\tau}(t)}{\partial y} = \frac12 \frac{\partial}{\partial y}(y \dot{s}_z + \dot{s}_z y) = \dot{s}_{z} ,
\end{align}
we can derive a conservation law for spin current:
\begin{align}
\frac{\partial S_{z}}{\partial t} + \frac{\partial j^{y}_{s}}{\partial y} = -\dot{s}_{z}.
\label{CL_Conserved}
\end{align}
The right-hand side represents spin relaxation due to the spin flip term in, e.g., the Rashba term. For a system where the total $s_z$ is conserved, the conservation law for spin is reinstated, as $\dot{s}_z = 0$. In such a case, the spin current exhibits a property akin to the charge current derived from charge conservation; the quantity of charge/spin accumulated on the material surface per unit time should be equivalent to the charge/spin current flowing in the bulk. However, given that spin is typically not conserved ($\dot{s}_z \neq 0$), there is no assurance that Eq.~(\ref{SC_Conserved}), the new definition of spin current, correlates with the spin accumulation on the material surface.

The SHC formulated using conserved spin current is given as \cite{Shitade2022}
\begin{align}
\sigma^{xy}_{s}&=-\frac{e}{\hbar}\sum_{n}\int\frac{d^2 k}{(2\pi)^2}[s^{x}_{n}\partial_{k_y}\epsilon_n f'(\epsilon^{\bm{k}}_n)+\epsilon^{xy\bm{k}}b_{n z\bm{k}}f(\epsilon^{\bm{k}}_n)],
\label{SHC_Conserved} \\
s^{x}_{n}&=-\frac{i}{2}[\braket{{u^{\bm{k}}_n}|s_z Q_{n}|{\partial_{k_{y}} u^{\bm{k}}_n}}-{\rm c.c.}], \\
\epsilon^{xy\bm{k}}b_{nz\bm{k}}&=\frac{i}{2}[\braket{\partial_{k_y} u^{\bm{k}}_{n}|Q_{n}\left[s_z Q_{n}+\braket{{u^{\bm{k}}_n}|s_z|{u^{\bm{k}}_n}}\right]|\partial_{k_y} u^{\bm{k}}_{n}}-{\rm c.c.}] \nonumber \\
&-\frac{i}{2}\sum_{m \neq n}\frac{\braket{{u^{\bm{k}}_n}|s_z|{u^{\bm{k}}_m}}\left[\braket{{u^{\bm{k}}_m}|(\partial_{k_ x}\epsilon_{n}+\hbar v_{x})Q_{n})|\partial_{k_{y}}{u^{\bm{k}}_n}}-(x \leftrightarrow y)\right]-{\rm c.c.}}{\epsilon^{\bm{k}}_{n} - \epsilon^{\bm{k}}_{m}},
\end{align}
where $Q_{n} = 1 - \ket{u^{\bm{k}}_n} \bra{u^{\bm{k}}_n}$. When a finite band-gap energy is present, the Fermi energy exists within the band gap, allowing us to neglect the first term in Eq. (\ref{SHC_Conserved}). To perform the numerical calculation of Eq. (\ref{SHC_Conserved}), we need to obtain the eigen wavefunctions and eigenvalues by diagonalizing the system Hamiltonian.

\subsection{Kane-Mele Model in Wavenumber Space}
\label{subsec2_3}
To calculate the SHC given by Eq. (\ref{Conventional_1}) or Eq. (\ref{SHC_Conserved}), we need information on the band energy $\epsilon^{\bm{k}}_{n}$ and Bloch wavefunction $\ket{{u^{\bm{k}}_n}}$ of the Kane-Mele model. The Hamiltonian of the Kane-Mele model in $\bm{k}$-space can be obtained throughout the following transformation:
\begin{align}
a^{\dagger}_{j\sigma} = \frac{1}{\sqrt{N}}\sum_{\bm{k}}a^{\dagger}_{\bm{k}\sigma}e^{-i \bm{k} \cdot {\bm{R}}^{j}_{A}}, \quad
b^{\dagger}_{j\sigma} = \frac{1}{\sqrt{N}}\sum_{\bm{k}}b^{\dagger}_{\bm{k}\sigma}e^{-i \bm{k} \cdot \bm{R}^{j}_{B}}.
\label{FT2}
\end{align}
Here, we define $\bm{k}$ as the Bloch wave vector and $\bm{R}^{j}_{A}$ ($\bm{R}^{j}_{B}$) as the position of the site on sublattices A (B) for the $j$-th unit cell. By applying these transformations to Eq. (\ref{RS_Hamiltonian}), we obtain the total Hamiltonian in $\bm{k}$-space in the form, 
\begin{align}
H &= H_{0} + H_{v} + H_{\rm SO} + H_{\rm R}, \label{HK} \\ 
H_{0} &= t_{0}\sum_{\bm{k}\sigma} (f(\bm{k})a_{\bm{k}\sigma}^\dagger b_{\bm{k}\sigma} +f^{*}(\bm{k}) b_{\bm{k}\sigma}^\dagger a_{\bm{k}\sigma}) , \\
H_{v} &=\lambda_{v}\sum_{\bm{k}\sigma}(a^{\dagger}_{\bm{k}\sigma} a_{\bm{k}\sigma} - b^{\dagger}_{\bm{k}\sigma} b_{\bm{k}\sigma}), \\
H_{\rm SO} &= 2\lambda_{SO}\sum_{\bm{k}\sigma s_z}{\rm{Im}}[F(\bm{k})]s_{z}(a^{\dagger}_{\bm{k}\sigma} a_{\bm{k}\sigma} - b^{\dagger}_{\bm{k}\sigma} b_{\bm{k}\sigma}), \\
H_{\rm R} &= \sum_{\bm{k}}A^{+}(\bm{k})a_{\bm{k}\uparrow}^{\dagger} b_{\bm{k}\downarrow} + A^{-}(\bm{k})a_{\bm{k}\downarrow}^{\dagger} b_{\bm{k}\uparrow} + {\rm{h.c.}}
\end{align}
Moreover, we define $f(\bm{k})=\sum_{j}e^{-i \bm{k}\cdot \bm{\delta}_{j}}$, $F(\bm{k}) = e^{-i \bm{k}\cdot (\bm{\delta}_{2} - \bm{\delta}_{3})} + e^{-i \bm{k}\cdot (\bm{\delta}_{1} - \bm{\delta}_{2})} + e^{-i \bm{k}\cdot (\bm{\delta}_{3} - \bm{\delta}_{1})}$, and $A^{\pm}(\bm{k}) = i \lambda_{\rm{R}} \sum_{j}[\delta^{y}_{j} \pm i \delta^{x}_{j}]e^{i \bm{k}\cdot \bm{\delta}_{j}}$. Diagonalizing Eq.~(\ref{HK}) enables us to obtain the band energy $\epsilon^{\bm{k}}_{n}$ and Bloch wavefunction $\ket{u^{\bm{k}}_{n}}$. By substituting these quantities into Eq.~(\ref{Conventional_1}) and Eq.~(\ref{SHC_Conserved}), we can calculate SHC based on the conventional and conserved spin currents. In the following subsection, we refer to the characteristics of the band structure of the Kane-Mele model as a function of $\lambda_{SO}$, $\lambda_{v}$, and $\lambda_{\rm{R}}$.

\subsection{Band structure of the Kane-Mele model}
\label{subsec2_4}

\begin{figure*}[hp]
\centering
\includegraphics[width=17.8cm]{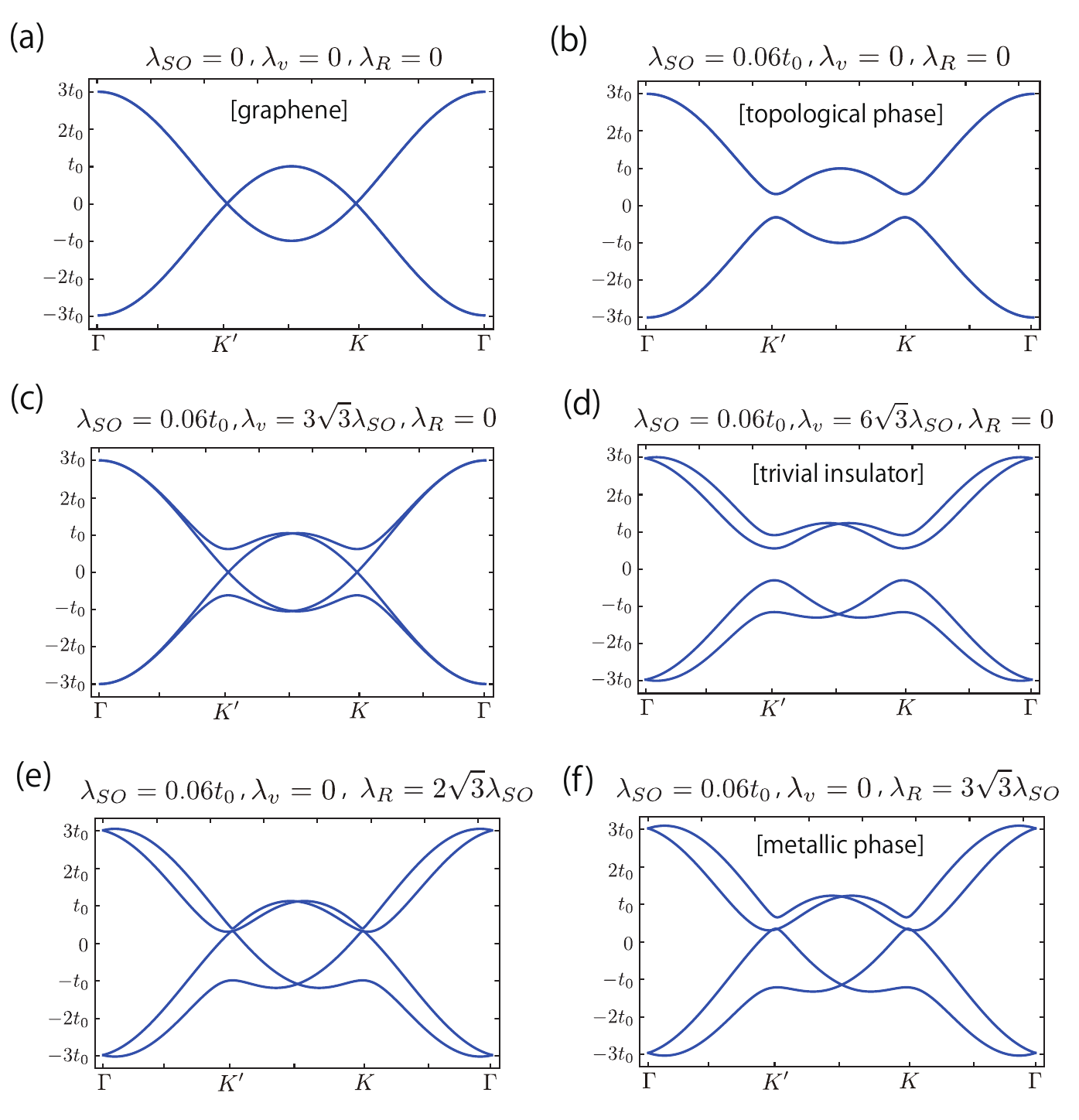}
\caption{(Color online) Band structures of the Kane-Mele model for different parameter sets $(\lambda_{\rm SO},\lambda_{v},\lambda_{\rm R})$: (a) $(0,0,0)$, (b) $(0.06 t_0,0,0)$, (c) $(0.06 t_0,3\sqrt{3}\lambda_{\rm SO},0)$, (d) $(0.06 t_0,6\sqrt{3}\lambda_{\rm SO},0)$, (e) $(0.06 t_0,0,2\sqrt{3}\lambda_{\rm SO})$, and (f) $(0.06 t_0,0,3\sqrt{3}\lambda_{\rm SO})$. Figures (c) and (e) represent the points where the band gap closes.}
\label{figS5}
\end{figure*}

The SHC is closely related to the existence of an energy gap in the bulk system's band structure. In this section, we show the band structure of the Kane-Mele model for various parameter sets.

In Fig.~\ref{figS5}(a), we depict the band energy of the Kane-Mele model for $\lambda_{\rm SO}= \lambda_{v}=\lambda_{\rm R}=0$. This band structure corresponds to graphene, where two Dirac cones emerge at the $K$ and $K'$ points. When the SOI is present in this system, the opening of a band gap leads to a topological phase in which the SHC is quantized, i.e., takes the universal value of $e/4\pi$ as shown in Fig.~\ref{figS5}(b). As the staggered potential $\lambda_{v}$ increases, the band-gap energy gradually decreases, but the SHC remains quantized until the band gap closes at $\lambda_{v}=3\sqrt{3}\lambda_{\rm SO}$ (Fig.~\ref{figS5}(c)). Further increases in $\lambda_{v}$ reintroduce a finite band gap, rendering the system a trivial insulator, wherein the SHC vanishes (Fig.~\ref{figS5}(d)). On the other hand, when $\lambda_{\rm R}$ increases from zero in the topological insulator phase ($\lambda_v=0$, $\lambda_{\rm SO} > 0$), the band gap is reduced and becomes zero at $\lambda_{\rm R}=2 \sqrt{3}\lambda_{\rm SO}$ (Fig.~\ref{figS5}(e)). In contrast to the case of increasing $\lambda_{v}$, further increases in $\lambda_{\rm R}$ do not open the energy gap and maintain the metallic properties of the system (Fig. \ref{figS5}(f)). Because the spin (or its component along a specific direction) is not conserved in the presence of the Rashba term, no clear prediction on the SHC is known. As such, the validity of the spin current concept becomes questionable, and further investigations involving numerical calculations are necessary.

\bibliography{SIreference}
\clearpage